\documentclass[prb,notitlepage,twocolumn,superscriptaddress]{revtex4}
\usepackage{amsmath,amssymb,color,graphicx,hyperref,amsthm,bm}

\usepackage{xcolor}
\usepackage{graphicx}
\usepackage{amsmath}
\usepackage{amssymb}
\usepackage{amsfonts}
\usepackage{bm}
\usepackage{hyperref}
\usepackage{enumerate}
\usepackage{longtable}

\definecolor{MS-color}{RGB}{128,0,128}
\definecolor{TW-color}{RGB}{0,130,128}
\definecolor{Error-color}{RGB}{250,50,50}

\graphicspath{{./images/}}

\begin{document}

\title{Skyrmions formation due to unconventional magnetic modes in anisotropic multi-band superconductors}
\author{Thomas Winyard}
\affiliation{Department of Physics, KTH-Royal Institute of Technology, Stockholm, SE-10691 Sweden}
\affiliation{School of Mathematics, University of Leeds, Leeds LS2 9JT, United Kingdom}
\author{Mihail  Silaev}
 \affiliation{Department of
Physics and Nanoscience Center, University of Jyv\"askyl\"a, P.O.
Box 35 (YFL), FI-40014 University of Jyv\"askyl\"a, Finland}
\author{Egor Babaev}
\affiliation{Department of Physics, KTH-Royal Institute of Technology, Stockholm, SE-10691 Sweden}

\begin{abstract}
Multiband superconductors have a sufficient number of degrees of freedom to allow topological excitations characterized by Skyrmionic topological invariants. In the most common, clean $s$-wave multiband, systems the 
interband   Josephson and magnetic couplings favours composite vortex solutions,   without a Skyrmionic topological charge. It was discussed recently that certain kinds of anisotropies lead to hybridisation of the interband phase difference (Leggett) mode with magnetic modes, dramatically changing the hydromagnetostatics of the system. Here we report this effect for a range of parameters   that substantially alter the nature of the topological excitations, leading to solutions characterized by a nontrivial  Skyrmionic topological charge. The solutions have a form of a coreless texture formed of spatially separated but bound excitations in each band, namely fractional vortices, each carrying a fraction of the flux quantum. We demonstrate that in this regime there is a rich spectrum of Skyrmion solutions, with various topological charges, that are robust with respect to changes of parameters of the system and present for a wide range of anisotropies.
\end{abstract}

\maketitle
\section{Introduction}
Superconducting materials are in general multiband \cite{Suhl,Moskalenko,tilley} and anisotropic \cite{ginzburg1952,kats1969some,tilley1965ginsburg}. One of the questions that was discussed recently is how multiple coherence lengths (that can have unconventional hierarchies in isotropic multicomponent theories \cite{Babaev.Speight:05,Silaev.Babaev:11,Carlstrom.Garaud.ea:11a,babaev2017type} ) change when anisotropy is included\cite{winyard2017}. Importantly, besides coherence lengths, when there are unequal annisotropies in multiple bands $|\psi_\alpha|e^{i\theta_\alpha}$ (where $\alpha$ is the band index) the electrodynamics is principally different from the London's massive vector field theory \cite{London1935} and its description requires several length scales   \cite{silaev2017non}.
The new electro-dynamical effects that arise include the phase difference mode $(\theta_\alpha-\theta_\beta)$ \cite{Leggett1966} hybridizing with the magnetic mode \cite{silaev2017non}. This leads to multiple magnetic field penetration lengths, which in turn allows magnetic field inversion for particular parameters. The additional penetration lengths affect the vortex solutions of such systems. The affects of such anisotropies on the multi-quanta vortex solutions in the full non-linear Ginzburg-Landau model has also recently been investigated\cite{silaev2017non,winyard2017}.

In this paper we demonstrate that, the unconventional hydromagnetostatics that were shown to stem from anisotropies, lead to substantial changes in the nature of the topological excitations in certain regimes. Namely we demonstrate that when anisotropies are sufficiently strong, the lowest energy topological excitations are Skyrmions, while composite vortices by contrast are not stable. Moreover we find that in the Skyrmionic regime, the spectrum of solutions is very rich, with stable high-topological-charge solutions.

Recently, many multiband superconductors have been discovered and most of them 
are anisotropic. Most importantly, they can have distinct anisotropies in different bands, which is crucial for the magnetic field hybridization with the interband phase difference mode. For example, in the two-band superconductor MgB$_2$ one of the Fermi surfaces is mostly isotropic, while the other is almost cylindrical, with the Fermi velocity anisotropy $v_{Fab}/v_{Fc} \approx 8.6 $ \cite{Brinkman2002}.
  The other example of a multiband anisotropic superconductor is Sr$_2$RuO$_4$\cite{Mackenzie2003,Huang2016}, characterized by strong London length anisotropy, although 
  the anisotropy of each band contribution is not known.
  The pronounced anisotropy is characteristic also to the iron-based superconductors. 
  For example, upper critical fields in the 122-compounds can differ several times \cite{Yuan2009, PhysRevB.89.134502} when applied along the c-axis or in the ab-plane. The most restricting condition for the observation of proposed effects comes from the requirement that the  interband pairing should be much smaller than the intraband ones. This condition is needed to have several distinct coherence lengths in the multiband model\cite{Silaev.Babaev:12}.  
  Besides that, all relevant modes should have approximately the same characteristic lengths for the coupling 
  of magnetic field and the order parameter degrees of freedom to be effective.   
  In particular that means the system has to be only weakly 
  type-II in the limit of temperatures very close to the critical one.

\subsection{The Model}
The Ginzburg-Landau free energy for a clean anisotropic $n$-band superconducting system is given by
\begin{eqnarray}
 F =\frac{1}{2}\int_{\mathbb{R}^3}\left\{ \sum^{n}_{\alpha = 1} \left(\gamma^{-1}_{ij\alpha}D_j\psi_{\alpha}\right)
 \left(\gamma^{-1}_{ik\alpha}  \overline{D_k\psi_{\alpha}}\right) + \boldsymbol{B}^2\right. \nonumber \\
 \left. +\sum^n_{\alpha  = 1}\frac{\Gamma_{\alpha}}{4}\left({\psi^0_\alpha}^2 - \left|\psi_\alpha\right|^2\right)^2
 - \sum_{\beta < \alpha}\eta_{\alpha \beta} \left|\psi_\alpha\right|\left|\psi_\beta\right| 
 \cos{\left(\theta_{\alpha\beta}\right)}\right\}
 \label{GL}
  \end{eqnarray}
 \noindent where $D_i = \partial_i + ieA_i$ is the covariant derivative and $\psi_\alpha = \left|\psi_\alpha\right|e^{i\theta_\alpha}$ represents the different superconducting components in different bands.
 The first two terms are kinetic and magnetic energy, while the third and fourth terms are the potential and Josephson terms, where $\theta_{\alpha\beta} = \theta_\alpha - \theta_\beta$ and $\gamma_\alpha$, ${\psi^0_\alpha}^2$ and $\eta_{\alpha\beta}$ are positive constants that determine the ground state of the system. Note that we will scale our system such that $e = 1$ and $\psi^0_\alpha = 1$ for simplicity.
   Greek indices will always be used to denote superconducting components and Latin indices will be spatial, with the summation principle applied for repeated Latin indices only. The anisotropy of the system is given by $\gamma_{ij\alpha}$ which represents a 2 dimensional diagonal matrix for each component,
 \begin{equation}
 \gamma_{ij\alpha} = \left(\begin{array}{ccc} \gamma_{x \alpha} & & 
 \\ & \gamma_{y \alpha} & 
 \\ & & \gamma_{z \alpha} \end{array} \right).
 \label{Eq:matrix}
 \end{equation}
In the potential terms ${\psi^0_\alpha}$, $\Gamma_{\alpha}$ and $\eta_{12}$ are positive real constants. The anisotropy does not necessarily have to have the above symmetry and indeed we will consider rotating the anisotropy axis later in the paper to see what effect it has on our solutions. The final term above is the Josephson inter-band coupling, where $\theta_{\alpha\beta} = \theta_\alpha - \theta_\beta$ is the inter-band phase difference between components $\alpha$ and $\beta$. We focus on the case where it breaks the  $U(1)^n$ symmetry to a $U(1)$ symmetry.

For a detailed discussion of the microscopic justification of such models in the clean case see 
\cite{Silaev.Babaev:12}.
Here we use dimensionless units, normalizing the
  length by the quantity which becomes proportional to the diverging coherence length in the limit $T\to T_c$, where $T$ is temperature and $T_c$ is the critical temperature. The order parameter is normalized to the quantity proportional to its bulk value. Hence the coefficients in Eq.(\ref{GL}) have numerical values $\sim 1$. This includes interband Josephson coupling $|\eta_{12}|\sim 1$, which means that in the non-normalized units its value is about the condensation energy, which becomes small close to $T_c$. Hence, our consideration is focused on the materials with rather small interband interaction. 
  The magnetic field is normalized by the thermodynamic critical field and the free energy density is normalized by the condensation energy at a given temperature. 
  The effective electric charge $e$ in general regulates the magnetic field localization length around vortices. 
  Although in the model we consider the case where magnetic field is hybridized with the order parameter modes and there is no single scale which determines its decay scale, in the limit $T\to T_c$ the constant $e$ is proportional to the Ginzburg-Landau parameter, attributed to the dominant component of the superconducting pairing fields. The effect which we obtain here takes place when $e$ take not particularly large values, thus we set $e=1$ throughout, but still in the type-II regime at temperatures very close to the critical one.

 In this paper we consider systems which are homogeneous along $z$-axis and therefore can be described by the 
two-dimensional models. 
Hence we assume that  magnetic field has only one component  $\boldsymbol{B} = \left(0,0,B\right)$ and 
the order parameter fields are defined on the orthogonal x-y plane. 
The elementary topological excitations of the model \ref{GL} are fractional vortices. A single fractional vortex is defined where only one of the phases winds by the minimal amount of $2\pi$: i.e. for a fractional vortex in the 1st component only, the integral around the core of the vortex $\oint \nabla \theta_1=2\pi, \oint \nabla \theta_{\beta \neq 1}=0$. If $\eta_{\alpha\beta}=0$, due to electromagnetic coupling between the condensates such a vortex carries only a fraction of the flux quantum and has logarithmically divergent energy (see detailed calculation e.g. in \cite{frac,npb} and for anisotripopic case in \cite{silaev2017non}). Only a bound state of fractional vortices where all fractions add to an integer have finite energy, which for equal electric charge coupling, constitutes an equal number of fractional vortices in all bands $\oint \nabla \theta_\alpha=2\pi N$, where $N$ is the winding number of the system $N \in \mathbb{Z}$ (also referred to as the quanta of the system).

Let us quickly review the energetics of vortex excitations in the isotropic London model. In the isotropic limit the solution is a logarithmically confined axially symmetric bound state of fractional vortices that have a common core around which all phases wind by the same amount $\oint \nabla \theta_\alpha=2\pi$. 
The reason why, in the isotropic limit of the model \ref{GL}, the electromagnetic coupling favours composite vortex solutions with overlapping cores, is that the gradients of the phase difference are decoupled from the vector potential, i.e. they represent counter-flow of the various components that involves no charge transfer. Such a neutral flow is energetically much more expensive than the supercurrent associated with the gradients of the phase sum, that are instead coupled to the vector potential. Then the solution with co-centered fractional vortices minimizes the energy.
The Josephson coupling $\eta_{\alpha\beta}\ne 0$, also favours co-axial cores of composite topological excitations in the isotropic case. In particular it leads to linear confinement for two well-separated vortices, due to the phase difference gradients becoming condensed into a string (coined a Josephson string) with crossection approximated by a sine-Gordon kink.

If we now return to the full non-linear isotropic Ginzburg-Landau model, there appears new contributions to the magnetic field energy that are proportional to cross-gradients of relative density and relative phase.
These contributions can be cast in the form of Skyrmionic topological charge
density \cite{Babaev.Faddeev.ea:02}. However the contribution of these terms is not sufficient to lead to the formation of Skyrmions, in the weak-coupling theory of a clean isotropic s-wave multiband superconductor.

Below we demonstrate that the situation is very different in the anisotropic case, due to the new effect: anisotropy-driven hybriziation of the Leggett mode and the magnetic mode \cite{silaev2017non}. We demonstrated that it becomes energetically favourable to split composite vortices into complicated extended bound states of fractional vortices. These in turn lead to the classification of the various vortex excitations by Skyrmionic topological invariants.

\subsection{Skyrmions}
As will be clear from the discussion below, the splitting of integer-flux vortices into fractional flux constituents leads to a new energetically conserved property: non-trivial Skyrmionic topological charge \cite{skyrme1962unified,skyrme1961non,Manton.Sutcliffe}. In the case of $n$ components the solutions are $\mathbb{C}P^{n-1}$ Skyrmions.

One can formulate the Skyrmionic topological invariant by combining our $n$ complex fields $\psi_\alpha$ into the complex $n$-vector $\Psi : \mathbb{R}^2 \rightarrow \mathbb{C}^n$. Note that if we restrict to configurations where there is no fractional vortex core overlap, namely $\Psi \neq 0$ anywhere, we can consider this to be the map $\Psi: \mathbb{R}^2 \rightarrow \mathbb{C}^n\backslash \left\{0\right\}$(where $\backslash$ denotes the relative complement). We also define the map $\pi : \mathbb{C}^n \backslash \left\{0\right\} \rightarrow \mathbb{C}P^{n-1}$ as the canonical projection to the complex line through the origin $0$ that contains the mapped point. The composition of these maps $\Phi = \pi \circ \Psi$ then takes each point $p\in\mathbb{R}^2$ in our physical space  to the equivalence class $\left[\Psi(p)\right] = \pi(\Psi(p)) \in \mathbb{C}P^{n-1}$. Namely, all points on the same line through the origin in the space $\mathbb{C}^n\backslash\left\{0\right\}$ are equivalent under the map $\pi$.

For physical reasons the field $\Phi$ must take its vacuum value on the boundary of the space (easily found to be a constant by substituting into the energy functional) as $\left|x\right| \rightarrow \infty$. The magnetic field can be defined by utilising the one-form
\begin{equation}
\nu = -Im\frac{X^\dagger dX}{\left|X\right|^2}
\end{equation}
where $X$ are the global coordinates on $\mathbb{C}^{n}\backslash \left\{0\right\}$. This leads to the following formulation for the magnetic field two-form and supercurrent,
\begin{eqnarray}
J &=& e \Psi^\dagger\Psi\left(e A - \Psi^\star \nu \right) \\
B = &dA& = \frac{1}{e}\left(d\left(\Psi^\star \nu\right) - \frac{1}{e}d\left(\frac{J}{\Psi^\dagger \Psi}\right)\right).\label{Eq:magneticfield}
\end{eqnarray}
where $\Psi^\star \nu$ is the pull-back of $\nu$ to $\mathbb{R}^2$ by the map $\Psi: \mathbb{R}^2 \rightarrow \mathbb{C}^k\backslash \left\{0\right\}$. Finally due to rewriting the exterior differential $d(\Psi^\star \nu) = \frac{1}{2} \Phi^\star \omega$, where $\omega$ is the Kahler form for the Fubini-Study metric on $\mathbb{C}P^{n-1}$, in case if there are no zeroes of the total density, the quantised magnetic flux can be derived from the above to be,
\begin{equation}
\int_{\mathbb{R}^2} B = \frac{1}{2e}\int_{\mathbb{R}^2} \Phi^\star \omega = \frac{2\pi}{e} Q(\Phi),
\end{equation}
which is determined by the homotopy class of the map $\Psi$ as $\omega$ is closed, which is equivalent to an integer value. For numerical work it is most convenient to be able to calculate $Q$ in integral form, hence it can be shown that
\begin{equation}
\Phi^\star \omega = \frac{2}{i\left|\Psi\right|^4}\left(\left|\Psi\right|^2 d\Psi^\dagger \wedge d\Psi + \Psi^\dagger d\Psi \wedge d\Psi^\dagger \Psi\right)
\end{equation}
which leads to a formulation for the Skyrmion charge \cite{Garaud.Carlstrom.ea:13},
\begin{equation}
Q\left(\Psi\right) = \int_{\mathbb{R}^2} \frac{i\epsilon_{ji}}{2\pi\left|\Psi\right|^4}\left[\left|\Psi\right|^2 \partial_i \Psi^\dagger \partial_j \Psi + \Psi^\dagger \partial_i \Psi \partial_j \Psi^\dagger \Psi\right] d^2 x.
\label{Eq:Q}
\end{equation}
This means we now have a set of distinct separate solution spaces, each characterised by a given Skyrmion charge or integer.

For the simplest   case in this article $n=2$, $Q\left(\Psi\right)$ gives the winding number of the map $\Phi : \mathbb{R}^2 \rightarrow S^2$, where the target space $\mathbb{C}P^1$ is identified with the 2-sphere $S^2$. For $n>2$ this is still an integer as required, however the map is no longer a sphere and the image of $Q$ is homologous to $Q$ copies of the generator of $H_2\left(\mathbb{C}P^{n-1}\right)$. 

Finally it is important to stress that this is not a standard (in the mathematical sense) topological charge and is not universally conserved for all parameters. The above maps all depend on the zero being removed from the target space of $\Psi$, hence if $\Psi$ vanishes then the topological arguments collapse and $Q = 0$. Hence the system has a finite rather than infinite potential barrier for changing the topological charge (or moving between the distinct solution spaces). That is, a configuration with a given magnetic flux can be deformed at finite energy cost to a configuration with lower topological charge by forcing the cores of the fractional vortices to coincide. Yet such ``reduced charge" configurations are not energetically stable for the regimes considered below and are excluded from the ground state. Moreover they are obviously entropically disfavoured at finite temperature and therefore,  in practise, the above quantities can be regarded as topological invariants. 

\section{Numerical Results}
Due to the highly nonlinear nature of the model the only way to   discover the lowest energy topological excitations is to consider accurate numerical simulations of various kinds. This is still a challenging problem due to the number of length scales that are involved in the interactions between vortices leading to increased chance of many local minima existing. All numerical solutions in this paper were found using the FreeFem++ library on a finite element space. A conjugate gradient flow method was applied to find the local minima from a given initial condition. Multiple initial configurations were used for any given solution, taking the form of perturbed spherically symmetric vortices either with higher winding number or well separated. We have simulated many more parameters than shown in this paper, confirming that the effects are present in a very wide range of parameters. The most representative values were chosen to display our results.
For all the simulations of isolated Skyrmions the numerical grid is much larger than the Skyrmion size, such that the Skyrmions experience no boundary interactions. Finally note that in this section and the following one the boundary conditions are such that $\boldsymbol{\nabla} \times \boldsymbol{A} = 0$ and $\boldsymbol{n}\cdot\boldsymbol{D}\psi_\alpha = 0$, where $\boldsymbol{n}$ is orthogonal to the boundary. Hence there is no external magnetic field being applied, this will be changed later in section IV. 

\subsection{Two Component $n=2$}
\begin{figure}[tb!]
 \includegraphics[width=1.0\linewidth]{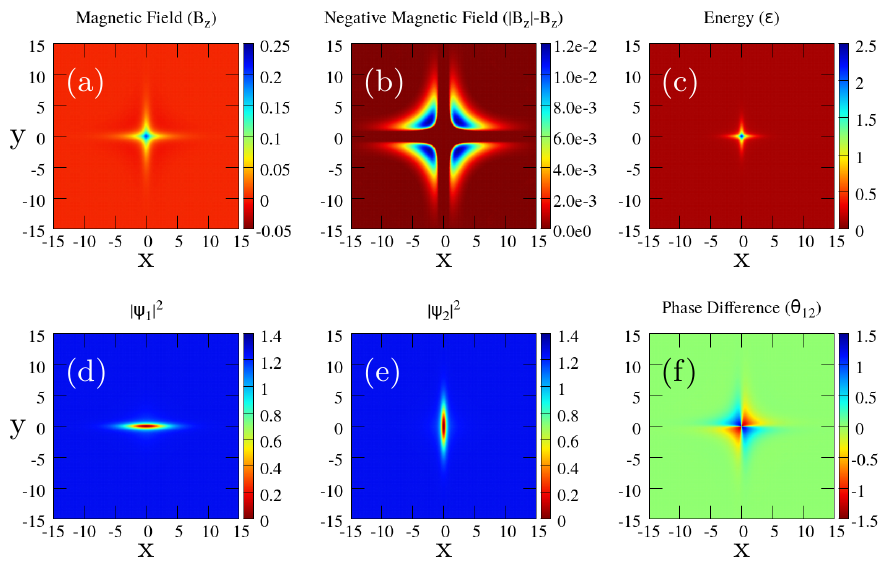}    
 \caption{\label{Fig:charge1} (Colour online)
 $N=1$, $Q=0$ single quanta numerical solution for anisotropy in both bands in opposite directions $\gamma_{x1}=\gamma_{y2}=2$, $\gamma_{y1} = \gamma_{x2} = 0.2$, $\eta_{12} = 0.5$ and $\Gamma_{1}=\Gamma_{2}=1$. The contour plots are (a) $B_z$ magnetic field (b) $\left|B_z\right| - B_z$ negative magnetic field (c) $\mathcal{E}$ energy density (d) $\left|\psi_1\right|^2$ (e)$\left|\psi_2\right|^2$ (f)$\theta_{12}$ phase difference.}
 \end{figure}

\begin{figure}[tb!]
   \includegraphics[width=1.0\linewidth]{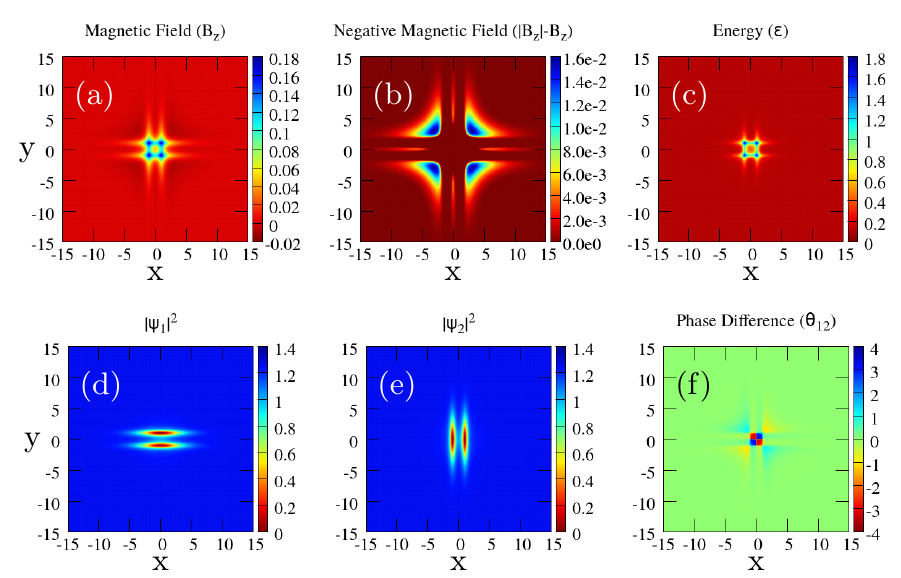}    
 \caption{\label{Fig:charge2} (Colour online)
 $N=2$, $Q=2$ two quanta numerical solution for anisotropy in both bands in opposite directions $\gamma_{x1}=\gamma_{y2}=2$, $\gamma_{y1} = \gamma_{x2} = 0.2$, $\eta_{12} = 0.5$ and $\Gamma_{1}=\Gamma_{2}=1$. The contour plots are (a) $B_z$ magnetic field (b) $\left|B_z\right| - B_z$ negative magnetic field (c) $\mathcal{E}$ energy density (d) $\left|\psi_1\right|^2$ (e)$\left|\psi_2\right|^2$ (f)$\theta_{12}$ phase difference.}
 \end{figure}

\begin{figure}[tb!]
  \includegraphics[width=1.0\linewidth]{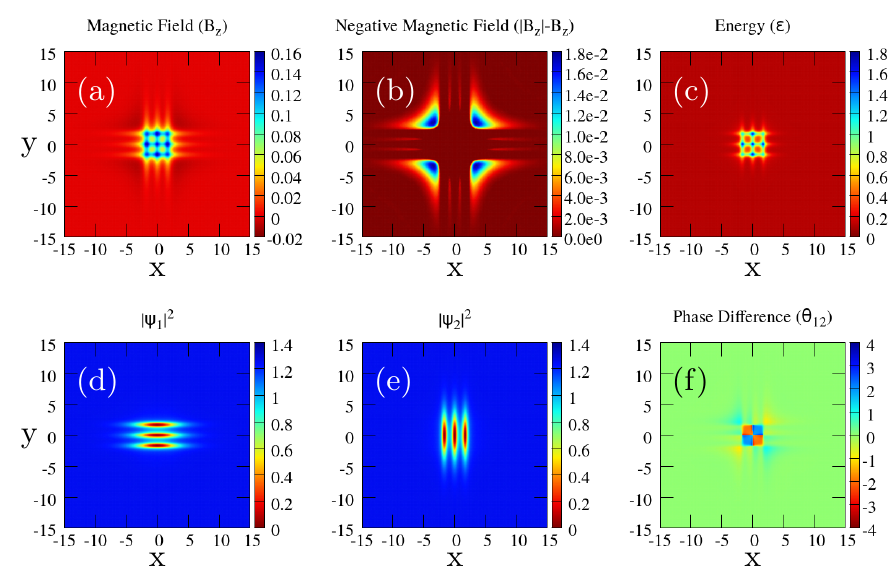}    
 \caption{\label{Fig:charge3} (Colour online)
 $N=3$, $Q=2$ three quanta numerical solution for anisotropy in both bands in opposite directions $\gamma_{x1}=\gamma_{y2}=2$, $\gamma_{y1} = \gamma_{x2} = 0.2$, $\eta_{12} = 0.5$ and $\Gamma_{1}=\Gamma_{2}=1$. The contour plots are (a) $B_z$ magnetic field (b) $\left|B_z\right| - B_z$ negative magnetic field (c) $\mathcal{E}$ energy density (d) $\left|\psi_1\right|^2$ (e)$\left|\psi_2\right|^2$ (f)$\theta_{12}$ phase difference. }
 \end{figure}

\begin{figure}[tb!]
 \includegraphics[width=1.0\linewidth]{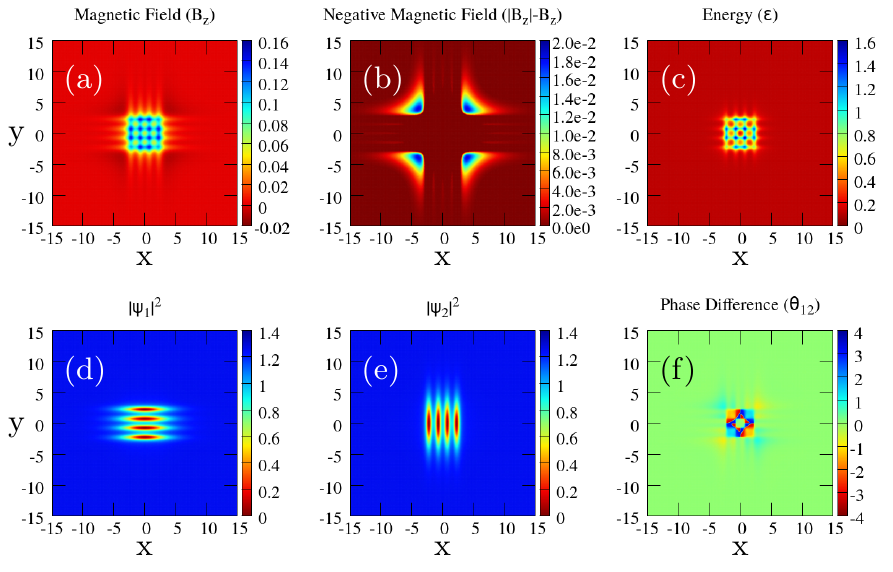}    
 \caption{\label{Fig:charge4} (Colour online)
 $N=4$, $Q=4$ four quanta numerical solution for anisotropy in both bands in opposite directions $\gamma_{x1}=\gamma_{y2}=2$, $\gamma_{y1} = \gamma_{x2} = 0.2$, $\eta_{12} = 0.5$ and $\Gamma_{1}=\Gamma_{2}=1$. The contour plots are (a) $B_z$ magnetic field (b) $\left|B_z\right| - B_z$ negative magnetic field (c) $\mathcal{E}$ energy density (d) $\left|\psi_1\right|^2$ (e)$\left|\psi_2\right|^2$ (f)$\theta_{12}$ phase difference. }
 \end{figure}

We start with the simplest case of considering the two band case ($n=2$). Note that for this band number the Skyrmion charge $Q$ is a winding number and has an intuitive form in terms of the $SU(2)$ generators, the Pauli matrices. We first define the projection vector,
\begin{equation}
\boldsymbol{n} = \frac{\Psi^\dagger \boldsymbol{\sigma} \Psi}{\Psi^\dagger \Psi}
\label{Eq:suproj}
\end{equation}
where $\boldsymbol{\sigma}$ is the vector of Pauli matrices. $\boldsymbol{n}$ can be thought of as a unit vector, representing the point on the target sphere $S^2$ by it's normal vector at that point. Hence for the texture $\boldsymbol{n}: \mathbb{R}^2 \rightarrow S^2$ to wind around the target sphere, all possible directions of $\boldsymbol{n}$ must exist in the physical space in a localised area. This intuitive projection leads to a simple formulation for the topological charge as,
\begin{equation}
Q(\boldsymbol{n}) = \frac{1}{4\pi}\int_{\mathbb{R}^2} \boldsymbol{n} \cdot \partial_x \boldsymbol{n} \times \partial_y \boldsymbol{n} \, dx dy.
\end{equation}
This quantity is zero for vortex solutions in the isotropic limit of the model.
For the solutions shown below, a numerical calculation of that topological charge gives an integer number with a numerically good accuracy.
We now consider taking similar anisotropies in each band but in opposite directions. This gives the energy functional a spatial $D_4$ symmetry, which is exhibited in the ($N=1$) single winding number solution shown in figure \ref{Fig:charge1}. Note that for the single quanta ($N=1$) case above there is no Skyrmion ($Q=0$) and hence no separation of the fractional vortices, which form a composite vortex with properties similar to those discussed in Ref.\onlinecite{silaev2017non,winyard2017}. However there is a nontrivial dependence of solutions on the number of flux quanta. If we then consider the two quanta ($N=2$) solution for these parameters, shown in figure \ref{Fig:charge2}, we see that we now have a bound state in the form of a Skyrmion solution with $Q=2$.
Observe that the fractional vortices are split, hence forming a Skyrmion solution. If we now consider the three quanta ($N=3$) solution in figure \ref{Fig:charge3} we notice that while a Skyrmion with $Q=2$ is formed, the Skyrmionic topological charge does not coincide with the number of flux quanta carried by that solution. This is because there is a composite vortex in the centre of the configuration. This leads to the conclusion that, while this trend continues, odd winding configurations have $Q=N-1$ with a single composite vortex in the centre and for even winding configurations $Q=N$. The evidence supporting that  prediction  comes from considering the next quanta solution $N=4$, $Q=4$ in figure \ref{Fig:charge4}.
 On both plots Fig.\ref{Fig:charge2} and Fig.\ref{Fig:charge3} one can see that the magnetic field behaves non-monotonically and has a small inverted tail far away from Skyrmion. This is shown in the plot of the negative magnetic field density ( $B_{neg} = \left|B_z\right| - B_z$ ) where $\left|\cdot\right|$ denotes the absolute value, hence $B_z>0 \implies B_{neg}=0$ and  $B_z\leq 0 \implies B_{neg}=2B_z$.
This is analogous to the field inversion around composite vortices in  multiband anisotropic superconductors
\cite{silaev2017non}. However, it is not the reason for the Skyrmion formation considered in the present paper. This can be seen by considering the magnetic field profiles in Figs.\ref{Fig:charge2}b and \ref{Fig:charge3}b, where it is clear that the field inversion takes place very far from Skyrmions. 
This emphases that the Skyrmion bound states are very different and occur at very different length scales than the vortex bound states that were found in different regimes in Ref. \onlinecite{silaev2017non}.
  
\begin{figure*}[tb!]
 \includegraphics[width=1.0\linewidth]{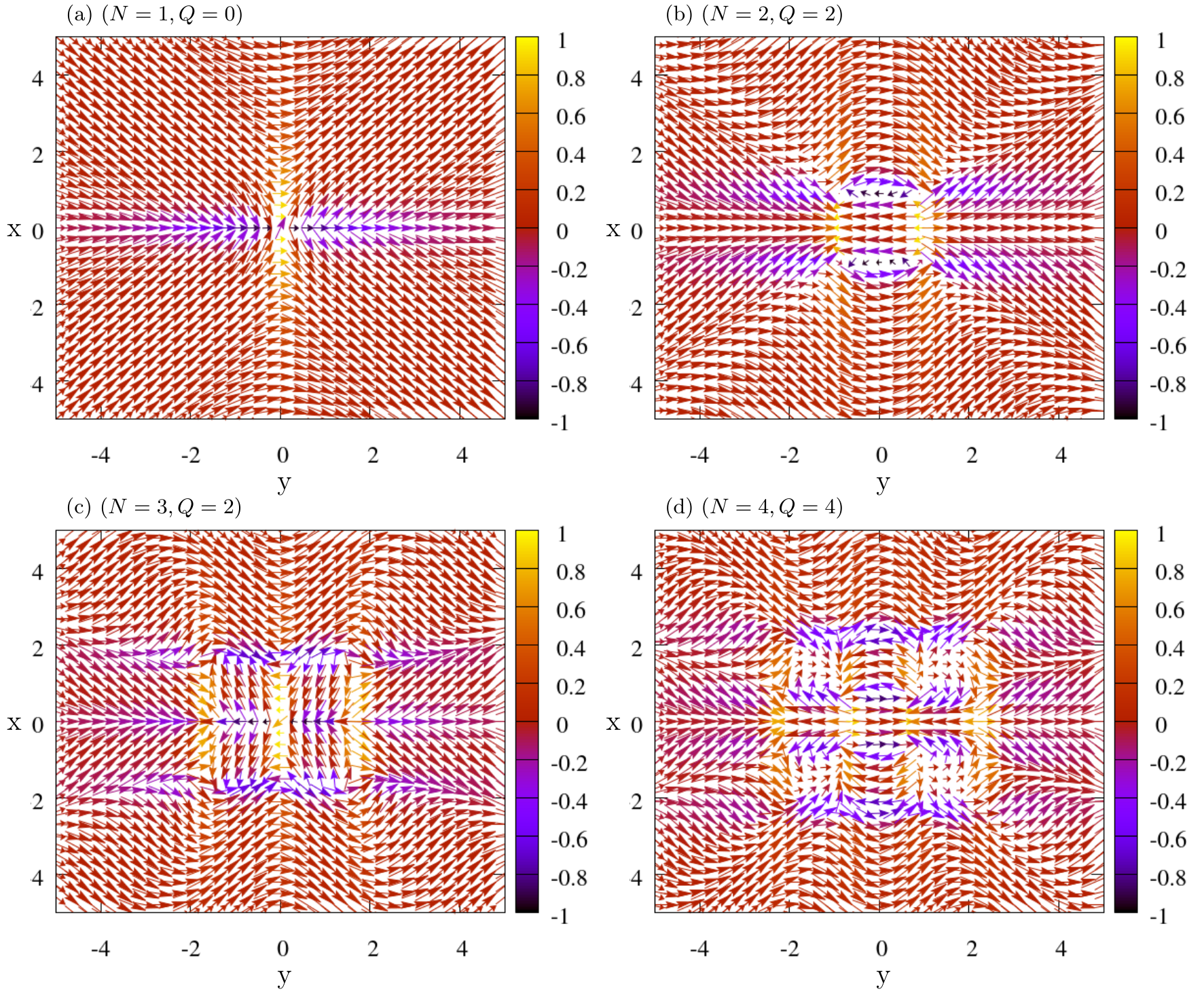}  
 \caption{\label{Fig:winding} (Colour online)
 Plots of the $SU(2)$ projection $\boldsymbol{n}$, given in equation \ref{Eq:suproj}, for $N=1-4$ one to four quanta numerical solutions for anisotropy in both bands in opposite directions $\gamma^{-1}_{x1}=\gamma^{-1}_{y2}=2$, $\gamma^{-1}_{y1} = \gamma^{-1}_{x2} = 0.2$, $\eta_{12} = 0.5$ and $\Gamma_{1}=\Gamma_{2}=1$. The arrows are coloured based on the value of $z$-component $n_z$.}
 \end{figure*}

 We have also plotted the spin $\boldsymbol{n}$ for the first four quanta solutions in figure \ref{Fig:winding}. The colour of the plots shows the $n_z$ component, where $n_z = 1$ (yellow) for $|\psi_1|^2 = 0$ and $n_z = -1$ (black) for $|\psi_2|^2 = 0$ which can be seen by substituting the zeroes into equation \ref{Eq:suproj}. These zeroes can be interpreted as the locations of the fractional vortex cores in the relevant condensates. $n_z = \pm 1$ can be thought of as the north/south pole of the target space, hence to cover the target space (and hence wind around it) they must be distinct and separate. Hence we can think of figure \ref{Fig:winding}(a) having $N= 1$ and $Q=0$ as bringing the north and south pole of the target space together in physical space (leading to $n_z$ being ill defined at that point in the physical space) and $n$ does not wind around the target $S^2$ space. However in figure \ref{Fig:winding} (b) with $N=2$ and $Q=2$ we can see the separate locations of the fractional vortices at the light and dark spots leading to winding for the map $n: \mathbb{R}^2 \rightarrow S^2$ covering the target space twice.

The obtained solutions clearly show that forces between fractional vortices are much more complicated in the anisotropic model, compared to the isotropic limit (outlined in the introduction). The physical origin of this is the strong anisotropy, which affects the strength of the coupling between the magnetic field and phase difference. This was shown in \cite{silaev2017non} and can be seen easiest by considering Amperes equation in the London model (taking the limit $\Gamma_\alpha \rightarrow \infty$ and hence assuming the magnitudes of the condensates take their vacuum value everywhere $\left|\psi_\alpha\right| = \psi_\alpha^0$). This leads to,
 \begin{align}\label{Eq:MagneticField}
 & {\bm B} = - \nabla\times \left( \hat\gamma_L^2 {\bm j} \right)
 + \\ \nonumber
 & \frac{1}{2}\sum_{\alpha>\beta}\nabla\times
 \left[ \hat\gamma_L^2 \left(\hat\gamma_{\alpha}^{-2} - \hat\gamma_{\beta}^{-2} \right) \nabla {\theta}_{\alpha\beta}
 \right],
 \end{align}
 where $ \hat\gamma_{\alpha}^{-2} = |\psi_\alpha|^2 \gamma_\alpha^{-2} $, where $\gamma_\alpha$ is the spatial matrix defined in \ref{Eq:matrix} and $\hat\gamma_L^2 = (\sum_\alpha \hat\gamma^{-2}_{\alpha})^{-1}$, where the spatial matrix indices are suppressed for $\hat \gamma$. It can easily be seen that the coefficient for the gradient of the phase difference is dependent on the strength of the anisotropy in the system.
 
The formation of Skyrmions is indeed a direct result of the introduction of anisotropy. This is supported numerically by considering the isotropic model with our chosen potential terms, for which Skyrmion solutions do not form. When a particularly strong anisotropy is introduced however they become energetically favourable solutions for degree (quanta) $N\geq 2$.  
The anisotropy creates hybridization of the phase difference and magnetic modes \cite{silaev2017non}.
That means that in the presence of magnetic field, the system creates phase difference gradients.
When the anisotropy is sufficiently strong it becomes energetically preferred to split integer vortices into fractional ones.
 In the examples we have plotted, we have considered anisotropy that is equal and opposite in terms of the $x$ and $y$ direction. This means the two different components want their fractional vortices to split and repel from a composite vortex along different axis. Ultimately this leads to a bound state of a Skyrmion as long as the anisotropy is strong enough.

We emphasise that solutions we observe in this paper, require strong anisotropy such that we enter a regime where the hybridization of the Leggett's (phase difference) mode with the magnetic mode becomes strong. This new mode introduces additional length scales and affects the interactions of the fractional vortices as shown in \cite{silaev2017non}. It should be noted that  the Skyrmion bound states are not the result of the field inversion effect  that leads to the
formation of the vortex bound states considered in  \cite{silaev2017non}. 
The Skyrmion splitting occurs on a shorter range and is strongly affected by non-linearities. This makes it hard to estimate analytically for which parameter values splitting will occur, beyond the requirement for strong hybridization of the phase difference and magnetic field. However it can be straightforwardly seen that the length scales are different for Skyrmion formation compared to the bound states formed by long-range interactions by considering the plots \ref{Fig:charge1},\ref{Fig:charge2},\ref{Fig:charge3} and \ref{Fig:charge4}, where it can be seen that the negative magnetic field occurs on a far longer range length scale than the fractional vortex splitting. The solutions are also different from 
bound states of vortices due to attractive density-density interactions that occur in a different regime \cite{winyard2018hierarchies}.


 The main point of the numerical solutions of this work is the demonstration of, in contrast to the isotropic case, under strong anisotropy, the interaction between fractional vortices changing from short-range attractive to short-range repulsive. This makes it energetically favourable to split fractional vortices into Skyrmions.

\begin{figure}[tb!]
  \includegraphics[width=1.0\linewidth]{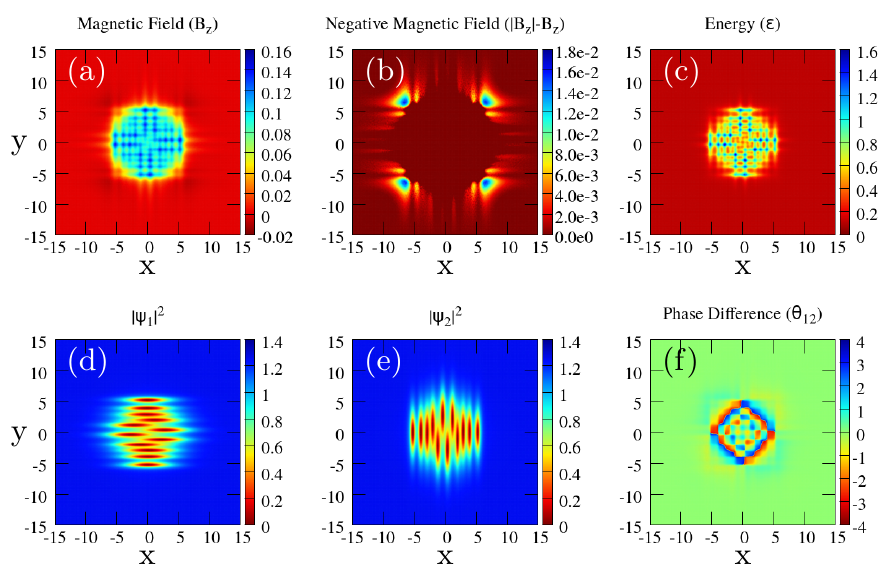}  
 \caption{\label{Fig:charge12} (Colour online)
 $N=12$ twelve quanta numerical solution for anisotropy in both bands in opposite directions $\gamma_{x1}=\gamma_{y2}=2$, $\gamma_{y1} = \gamma_{x2} = 0.2$, $\eta_{12} = 0.5$ and $\Gamma_{1}=\Gamma_{2}=1$. The contour plots are (a) $B_z$ magnetic field (b) $\left|B_z\right| - B_z$ negative magnetic field (c) $\mathcal{E}$ energy density (d) $\left|\psi_1\right|^2$ (e)$\left|\psi_2\right|^2$ (f)$\theta_{12}$ phase difference. }
 \end{figure}

It is also interesting how the solutions change with further increased topological charge.
There should be nontrivial scaling with increased flux quanta since the larger separation
of fractional vortices should result in the appearance of a linear energy penalty from the Josephson term. 
Thus we are interested in the high quanta solutions, the $N=12$ twelve quanta solution of which is plotted in figure \ref{Fig:charge12}. If we start with an initial configuration similar to what you may expect, of fractional vortices separated in the $x$/$y$ direction, like for small $N$ solutions,  it will collapse into the plotted solution, with fractional vortices breaking out of the line. 

\subsection{CP2 Skyrmions in three band model}
We will briefly touch on higher component solutions here, specifically $n=3$. Skyrmions still exist in this regime but are characterized by a $CP^2$ topological invariant (see equation \ref{Eq:Q}). This topological invariant $Q$ is still an integer, but cannot be interpreted as a winding number around a target sphere any more.

\begin{figure}[tb!]
  \includegraphics[width=1.0\linewidth]{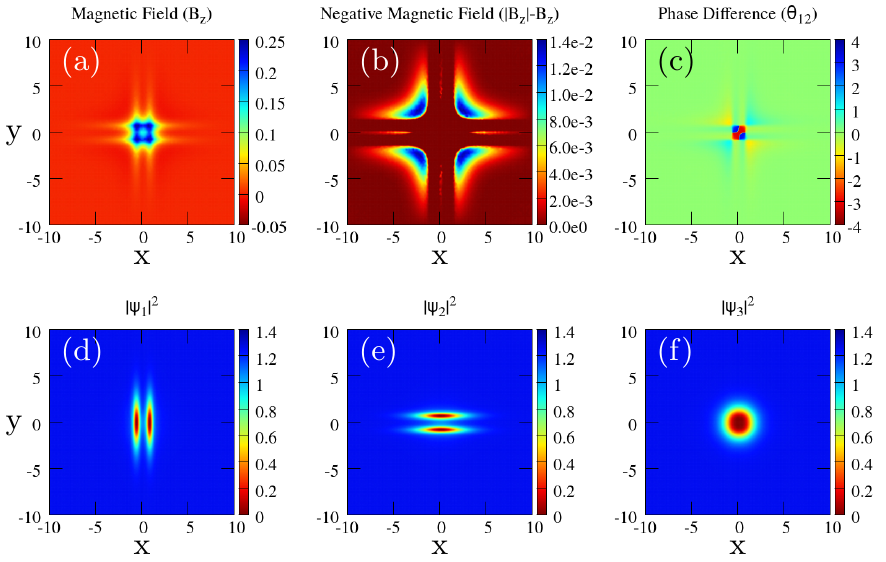}  
 \caption{\label{Fig:3comstan} (Colour online)
 $N=2$, $Q=2$ two quanta numerical solution for anisotropy in two bands in opposite directions and no anisotropy in the third $\gamma_{x1}^{-1}=\gamma_{y2}^{-1}=2$, $\gamma_{y1}^{-1} = \gamma_{x2}^{-1} = 0.2$, $\gamma_{3x}^{-1} = \gamma_{3y}^{-1}=1$, $\eta_{12} = \eta_{23} = \eta_{13} = 0.5$ and $\Gamma_{1}=\Gamma_{2}=\Gamma_{3}=0.5$. The contour plots are (a) $B_z$ magnetic field (b) $\left|B_z\right| - B_z$ negative magnetic field (c) $\mathcal{E}$ energy density (d) $\left|\psi_1\right|^2$ (e)$\left|\psi_2\right|^2$ (f)$\theta_{12}$ phase difference. }
 \end{figure}

We first present adding a band with no anisotropy, presented in figure \ref{Fig:3comstan}. We are interested in the parameters $\gamma_{1x}^{-1} = \gamma_{2y}^{-1} = 2$, $\gamma_{1y}^{-1} = \gamma_{2x}^{-1} = 0.2$ similar to previous simulations, while $\gamma_{3x}^{-1} = \gamma_{3y}^{-1} = 1$, with the parameters $\Gamma_1 = \Gamma_2 = 0.5$ and $\eta_{12} = \eta_{23} = \eta_{13} = 0.5$ such that the third component zeroes attract to form a higher winding fractional vortex at the centre of the Skyrmion, surrounded by the fractional vortices in the other components as shown in figure \ref{Fig:3comstan}. 

\begin{figure}[tb!]
  \includegraphics[width=1.0\linewidth]{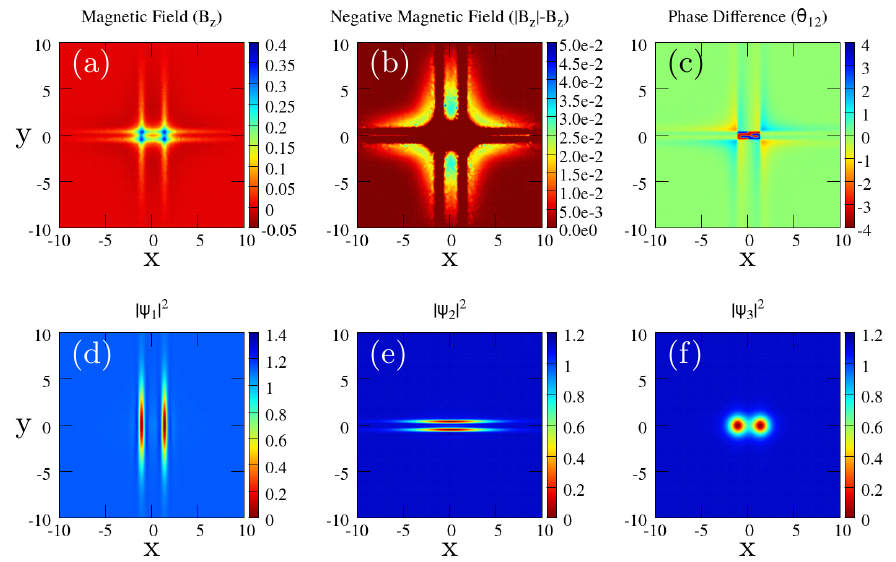}   
  \caption{\label{Fig:3comalt} (Colour online)
 $N=2$, $Q=2$ two quanta numerical solution for anisotropy in two bands in opposite directions and no anisotropy in the third $\gamma_{x1}^{-1}=\gamma_{y2}^{-1}=2$, $\gamma_{y1}^{-1} = \gamma_{x2}^{-1} = 0.2$, $\gamma_{3x}^{-1} = \gamma_{3y}^{-1}=1$, $\eta_{12} = \eta_{23} = \eta_{13} = 0.5$ and $\Gamma_{1}=\Gamma_{2}=\Gamma_{3}=2.0$. The contour plots are (a) $B_z$ magnetic field (b) $\left|B_z\right| - B_z$ negative magnetic field (c) $\mathcal{E}$ energy density (d) $\left|\psi_1\right|^2$ (e)$\left|\psi_2\right|^2$ (f)$\theta_{12}$ phase difference.  }
 \end{figure}

A different solution is possible if the vortices in the third band are also caused to split. This happens for the cases where the potential is stronger in the third band and hence the fractional vortices in this band repel each other in a stronger fashion, alternately it happens when there is anisotropy in the band to cause the vortices to repel in a particular direction. This leads to the fractional vortices splitting in this band and attempting to form a composite vortex with the other bands (in particular the band that is closest in anisotropy). This leads to the results plotted in figure \ref{Fig:3comalt} where we observe overlapping fractional vortices. This causes it to be more energetically favourable for the vortices to be at a higher separation in this particular direction (the type-2 nature of the 3rd band pushing the composite vortices apart) due to the higher magnetic field compared with the orthogonal direction giving a warped shape to the magnetic field profile.

We find therefore that Skyrmions do exist and are quite stable in the three component generalization and the trend should persist with increased number of bands.

\section{Rotated Anisotropies} 

In section II we discussed the basic model with anisotropies entering the model 
in the form of purely diagonal matrices $\gamma_\alpha$. This can be thought of as the anisotropies being orthogonal to each other. In this section we inspect 
whether or not this effect is related to one particular kind of anisotropy and how robust the Skyrmionic solutions are to altering this anisotropic symmetry. To that end we  consider the following extended model: the slight extension by applying rotations, independently to each band. Hence in 2-dimensions,
 \begin{equation}
 \gamma_{ij\alpha} = \left(\begin{array}{cc} \cos{\varphi_\alpha} & -\sin{\varphi_\alpha}
 \\ \sin{\varphi_\alpha}& \cos{\varphi_\alpha} \end{array}\right)  \left(\begin{array}{cc} \gamma_{x \alpha} &
 \\ & \gamma_{y \alpha} \end{array}\right).
 \end{equation}
Where $\varphi_\alpha$ are now parameters of the model, giving the rotation angle of the band $\alpha$. This leads to cross terms between $D_x$ and $D_y$, which could normally be removed should we be in the isotropic case, as the fields could be rewritten as a linear combination, cancelling the cross terms.

\begin{figure}[tb!]
   \includegraphics[width=1.0\linewidth]{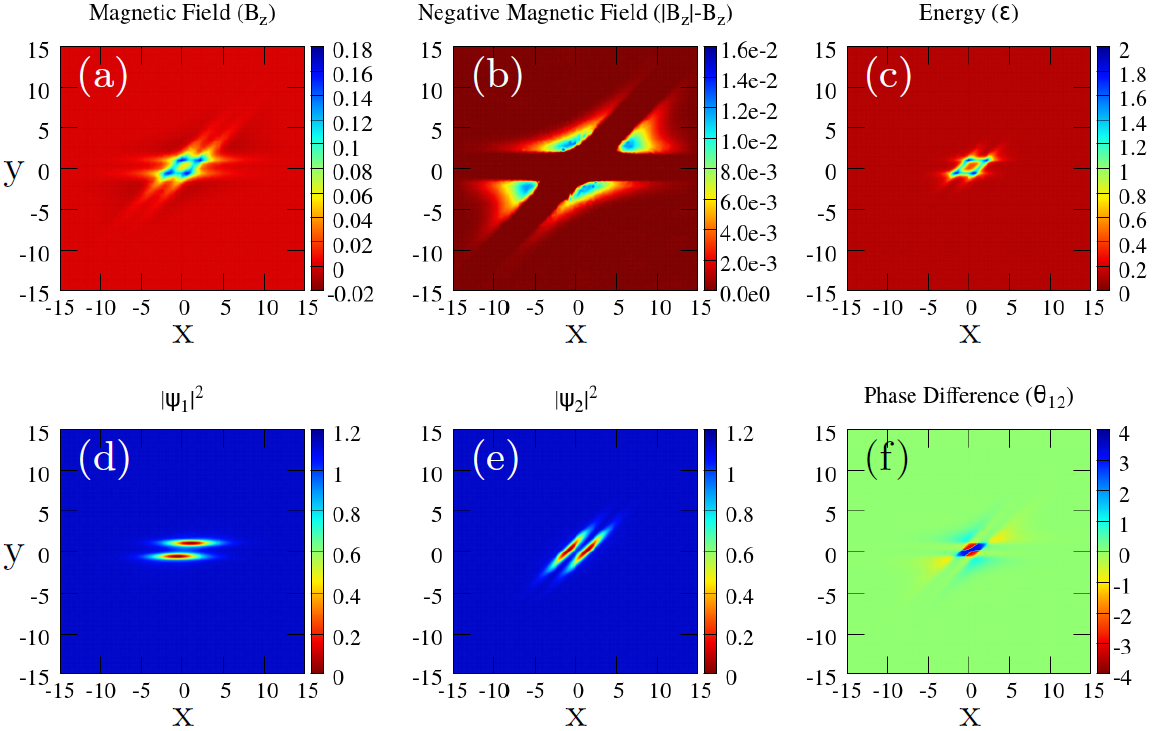} 
 \caption{\label{Fig:2comrot} (Colour online)
 $N=2$ two quanta numerical solution for anisotropy rotated differently in each band $\gamma_{x1}=\gamma_{x2}=2$, $\gamma_{y1} = \gamma_{y2} = 0.2$, $\varphi_1 = 0$, $\varphi_2 = \pi/4$, $\eta_{12} = 0.5$ and $\Gamma_{1}=\Gamma_{2}=2$. The contour plots are (a) $B_z$ magnetic field (b) $\left|B_z\right| - B_z$ negative magnetic field (c) $\mathcal{E}$ energy density (d) $\left|\psi_1\right|^2$ (e)$\left|\psi_2\right|^2$ (f)$\theta_{12}$ phase difference. }
 \end{figure}

The symmetry of the model will now heavily depend on the choice of $\varphi_\alpha$ and will not have the familiar four-fold symmetry from above. In figure \ref{Fig:2comrot} we have plotted the effect of rotating the previous model with $\varphi_1 = 0$, $\varphi_2 = \pi/4$. Here we see the loss of the $D_4$ symmetry we previously had, however the Skyrmion solutions are stable and the familiar Skyrmion structure remains. This means the Skyrmions are not reliant on the spatial symmetry of the anisotropies and should exist for all choices of $\varphi_\alpha$. 

\begin{figure*}[tb!]
   \includegraphics[width=1.0\linewidth]{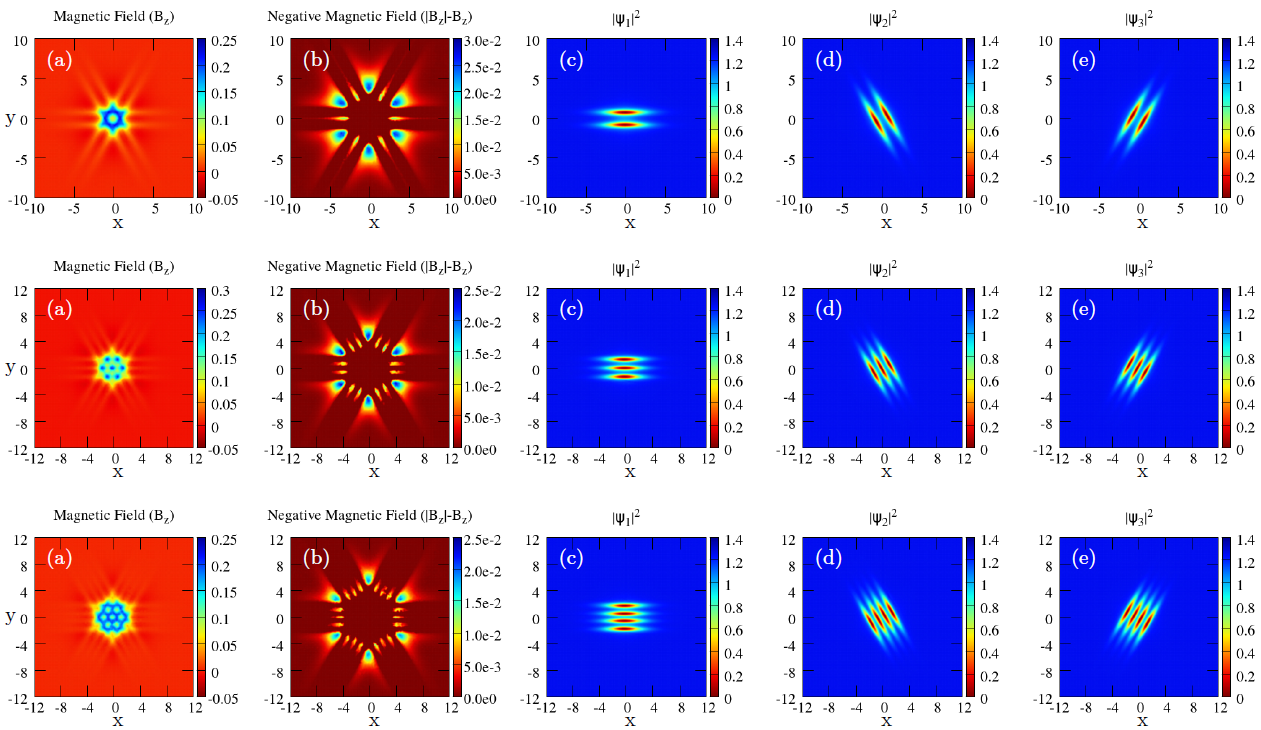} 
 \caption{\label{Fig:3comrot} (Colour online)
 $N=2,3,4$ quanta numerical solution for similar anisotropy in all the bands but with rotations $\varphi_1 = 0$, $\varphi_2 = 2\pi/3$, $\varphi_3 = 4\pi/3$ $\gamma_{x1}=\gamma_{x2}=\gamma_{x3} = 2$, $\gamma_{y1} = \gamma_{y2} = \gamma_{y3} = 0.2$, $\eta_{\alpha\beta} = 0.5$ and $\Gamma_{1}=\Gamma_{2}=\Gamma_{3}=2$. The contour plots are (a) $B_z$ magnetic field (b) $\left|B_z\right| - B_z$ negative magnetic field (c) $\left|\psi_1\right|^2$ (d)$\left|\psi_2\right|^2$ (e)$\left|\psi_3\right|^3$. }
 \end{figure*}

Further extension of the anisotropies considered in \cite{silaev2017non} are now possible. For example a $D_6$ dihedral or hexagonal symmetry can be formed in the 3 component model for $\varphi_1 = 0$, $\varphi_2 = 2\pi/3$ and $\varphi_3 = 4\pi/3$ and equal anisotropy in each band $\gamma_{1x} = \gamma_{2x} = \gamma_{3x}$, $\gamma_{1y} = \gamma_{2y} = \gamma_{3y}$ as plotted in figure \ref{Fig:3comrot}.

\section{Magnetization}
The above sections considered isolated Skyrmion solutions in the absence of an external magnetic field. We are now interested in how Skyrmions act when entering into a magnetised sample. To model the magnetization of a finite domain or sample we must introduce the external field $H$. 
 
Hence we are now modelling the free energy $F_{mag} = F - 2\int_{\mathbb{R}^2} H\cdot B \; d^2 x$. 
Note that this doesn't effect the field equations in the bulk of the theory, as the additions are constants (due to the integral of the magnetic field density being fixed through the topology of the map). It does however have an effect on the boundary conditions of the problem, such that we now have $\boldsymbol{\nabla}\times \boldsymbol{A} = H$ on the boundary of our space, as well as the other conditions previously used.
If we then slowly increase the external field value in steps of $10^{-2}$ we can simulate the turning up of an external field and the subsequent magnetization of the theory over our finite domain.

\begin{figure*}[tb!]
  \includegraphics[width=1.0\linewidth]{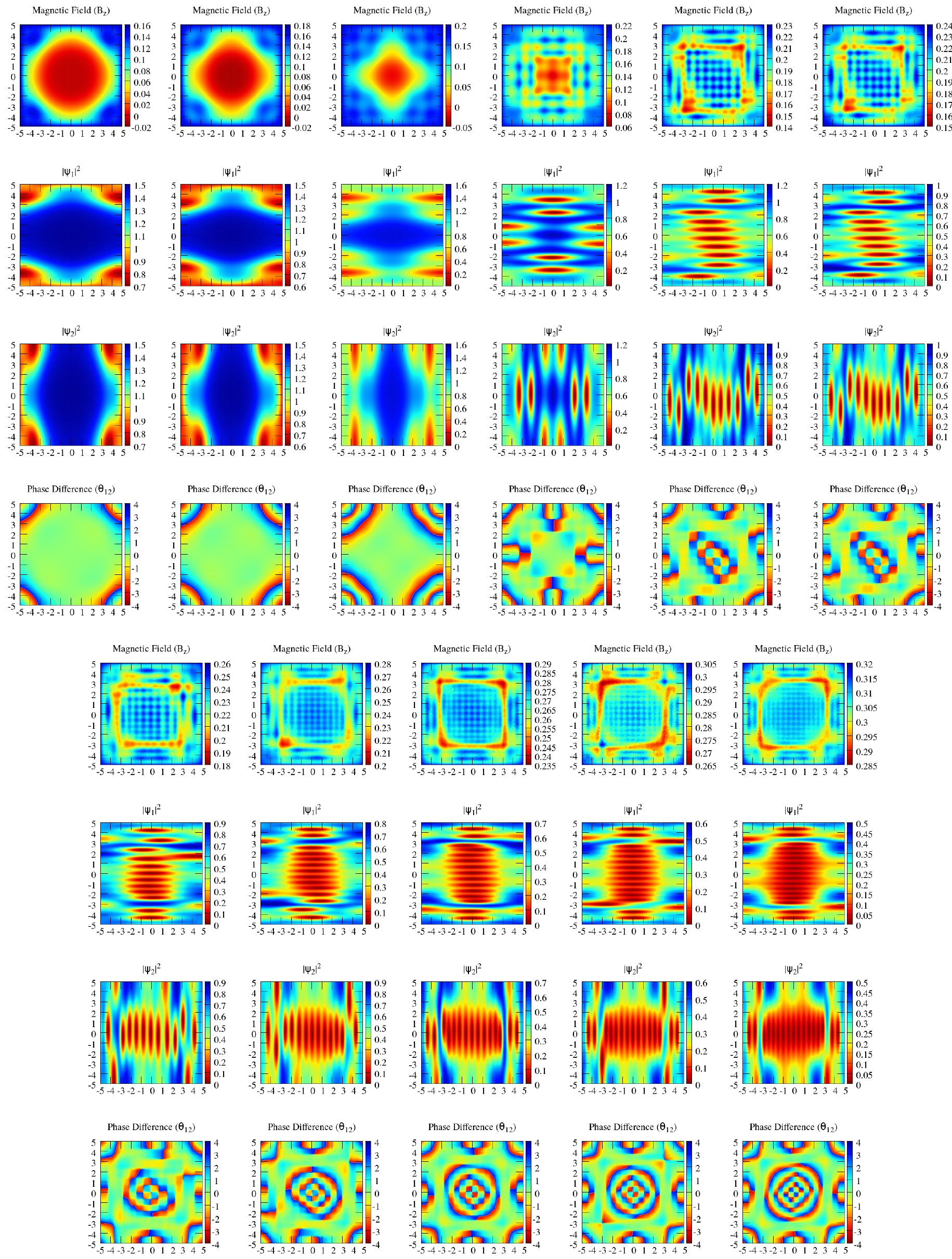}   
 \caption{\label{Fig:mag1} (Colour online)
 Successive plots of magnetising a sample with increasing external field $H$ running from left to right with parameters  $\gamma_{x1}=\gamma_{y2}= 2$, $\gamma_{y1} = \gamma_{x2} = 0.2$, $\eta_{\alpha\beta} = 0.5$ and $\Gamma_{1}=\Gamma_{2}=1$. The different quantities that are plotted from top to bottom are (a) $B_z$ magnetic field (b) $\left|\psi_1\right|^2$ (c)$\left|\psi_2\right|^2$ (d) $\theta_{12}$ phase difference. One an see that the cores of fractional vortices are not superimposed and thus the solution forming in an external field represents a a large-charge Skyrmion. }
 \end{figure*}

We will only consider the 2 band system here as the 3 band case can be extrapolated from this. We start with considering the parameters that we have considered in the previous sections in figure \ref{Fig:mag1}. We observe two chains of fractional vortices forming in alternate directions for each condensate. Hence we get Skyrmion solutions, similar to those considered in the mult-quanta solutions in section II.  While for an isolated solutions, the chain pattern tends to fracture, as with the $N=12$ solution plotted in figure \ref{Fig:charge12}, it doesn't happen in applied external field here.
This is because of the finite size of the system and interaction with the Meissner current near boundaries increasing the stability of this form of solution. In this situation the system tries to form a large Skyrmion that minimizes intervortex forces and interaction with the boundaries.

\begin{figure*}[tb!]
 \includegraphics[width=1.0\linewidth]{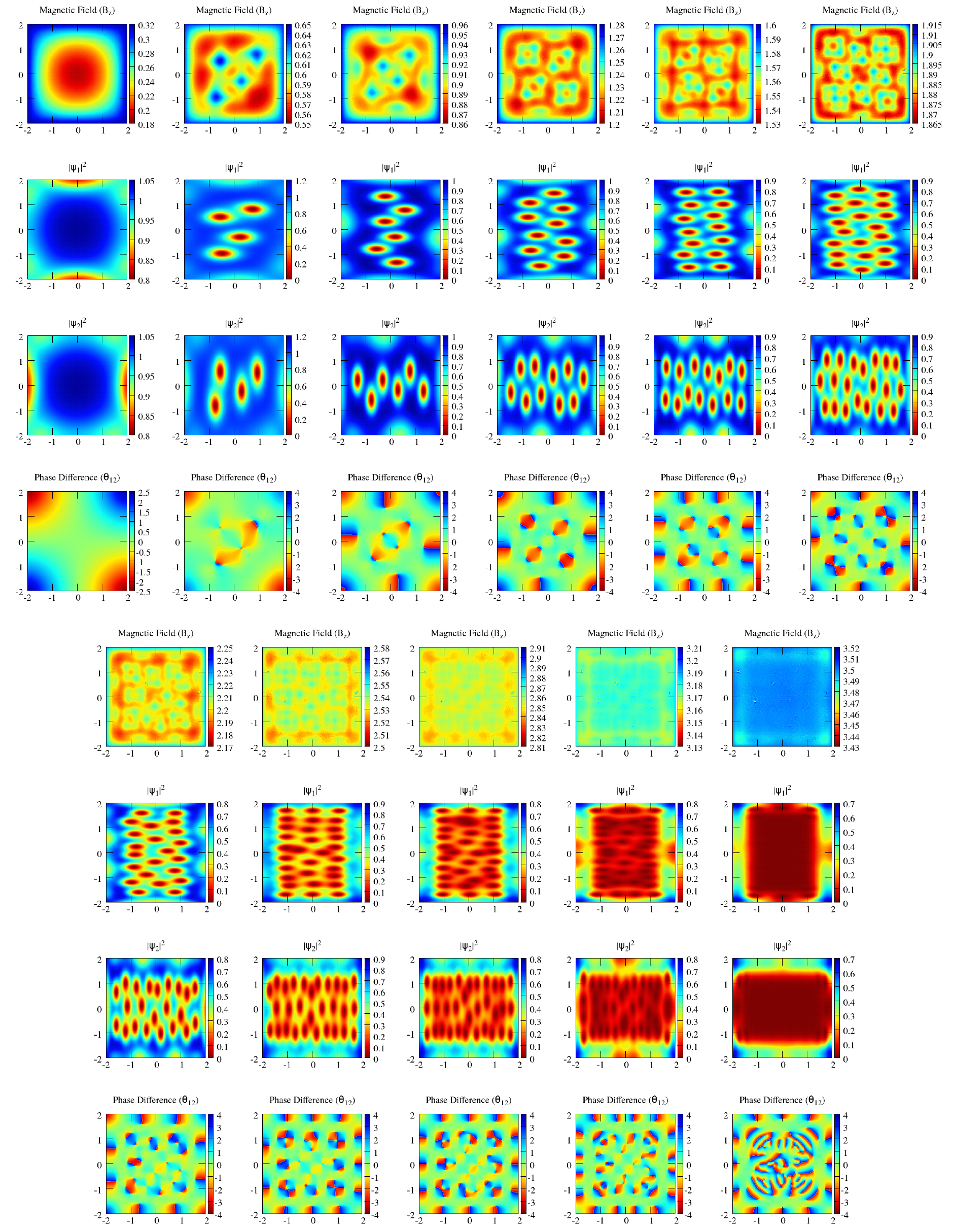}   
 \caption{\label{Fig:mag2} (Colour online) Successive plots of magnetising a sample with increasing external field $H$ running from left to right with parameters  $\gamma_{x1}=\gamma_{y2}= 0.8$, $\gamma_{y1} = \gamma_{x2} = 0.3$, $\eta_{\alpha\beta} = 0.5$ and $\Gamma_{1}=\Gamma_{2}=10$. The different quantities that are plotted from top to bottom are (a) $B_z$ magnetic field (b) $\left|\psi_1\right|^2$ (c)$\left|\psi_2\right|^2$ (d) $\theta_{12}$ phase difference. The length scale of the negative magnetic field mode which mediates the global minima bound states for the multi-quanta solutions is too long range to have a significant effect on this domain. Hence the rest of the hierarchy is important and this leads to Skyrmion formation on this short length scale.  }
 \end{figure*}

One of the key properties of the magnetization is the potential barrier to vortex entry being different for the individual boundaries due to the anisotropy. Namely, for the boundaries in figures \ref{Fig:mag1} and \ref{Fig:mag2}, the boundaries in the $x$ and $y$ directions will have different potential barriers for vortex entry for each condensate. In these examples the barrier is lower on different boundaries for each of the condensates. Hence this leads to fractional vortices entering into the domain on different boundaries for each condensate and hence are completely separated, stabilised by boundary interactions. Once they do enter into the sample, they enter as a pair of fractional vortices from orthogonal directions. Once in the bulk (together with a fractional vortex from the orthogonal direction) they are pulled into the chain solution emanating from the centre of the space. Due to this peculiar form for the energy potential barrier, we see fractional vortices entering predominantly from the corners of the space. This helps to break the chain apart at this point as the fractional vortices on the boundary and those already in the chain do not want to be aligned in the direction orthogonal to the chain length. This points towards the geometry of a domain heavily affecting the ease with which vortices can enter into the system with strong anisotropy present.

\begin{figure}[tb!]
   \includegraphics[width=1.0\linewidth]{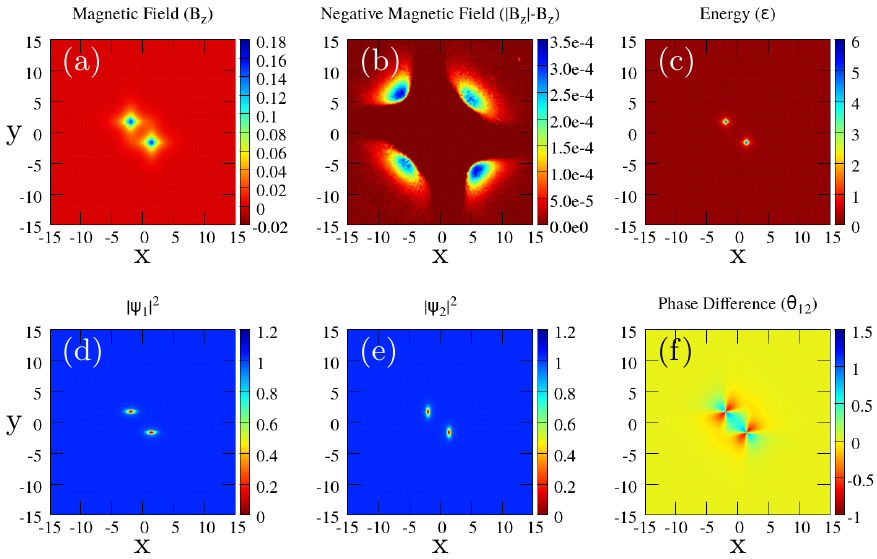}   
 \caption{\label{Fig:noskyrme} (Colour online)
 $N=2$ two quanta numerical solution with no Skyrmion solutions $Q=0$ for anisotropy in both bands in opposite directions $\gamma_{x1}^{-1}=\gamma_{y2}^{-1}=0.8$, $\gamma_{y1}^{-1} = \gamma_{x2}^{-1} = 0.3$,
  $\eta_{12} = 0.5$
   and $\Gamma_{1}=\Gamma_{2}=10$. The contour plots are (a) $B_z$ magnetic field (b) $\left|B_z\right| - B_z$ negative magnetic field (c) $\mathcal{E}$ energy density (d) $\left|\psi_1\right|^2$ (e)$\left|\psi_2\right|^2$ (f)$\theta_{12}$ phase difference.}
 \end{figure}

If we now consider a set of parameters that don't produce isolated Skyrmion multi-quanta states, a peculiar situation still occurs. In figure \ref{Fig:mag2}, we have plotted the magnetization of the parameters $\Gamma_1 = \Gamma_2 = 10$, $\gamma_{1x}^{-1} = \gamma_{2y}^{-1} = 0.8$, $\gamma_{1y}^{-1}=\gamma_{2x}^{-1} = 0.3$ and $\eta_{12} = 0.7$, 
 which do not produce Skyrmions for it's multi-quanta states, which can be seen in figure \ref{Fig:noskyrme} for it's two quanta state. We note that the minimal energy multi-quanta solutions for this systems are similar to those found in \cite{silaev2017non}, with the composite vortices positioning themselves in the negative magnetic field pocket of the other. This means that the form of this multiquanta solution and importantly the separation distance of the composite vortices that form it, is mediated by the second penetration length of the magnetic field which is longer range than the other length scales. In figure \ref{Fig:mag2}, we have considered the magnetisation of a sample that is smaller than this negative magnetic field length scale, leading to this penetration length having a minimal effect. This leads to very different behaviour and we see Skyrmions beginning to form as different intervortex and vortex-boundary forces dominate at these shorter ranges.

Initially the magnetisation in Fig.\ref{Fig:mag2} produces similar results to those in the previous case with fractional vortices entering from orthogonal directions due to differing energy barriers. However this time the anisotropy is smaller and the chain form breaks easily as in Fig.\ref{Fig:charge12}. Eventually the magnetic field becomes strong enough that the condensates begin to blend into a saturated domain. However due to the markedly different interactions with the boundaries in different directions, the edges of these domains do not correspond in the different condensates.

This means there are two different forms of solution, dependent on the length scale of the vortex separation. We predict that for a larger space one would observe the vortices in a distinct pattern similar to those shown in Ref. \onlinecite{silaev2017non} for the multi-quanta solutions, mediated by the long range negative magnetic field. However as the external field is increased and hence the density of vortices increases, there will be a shift in the solution as the repulsive part of the magnetic field dominates at the smaller length scale. This would lead to fractional vortex separation, similar to the form shown in Fig.\ref{Fig:mag2}. 
This suggests that when considering length scales that mediate very different multi-quanta vortex solutions, there will be a transition in the form of the bound states as the vortex separation changes it's scale from one of the length scales to the other. Namely, if the external magnetic field is increased, causing the magnetic field density to increase, the shorter length scales will start to dominate and the bound states will change.

{ It is important to note that these structures cannot be interpreted as a superposition of two single-component anisotropic lattices. This is because the components are strongly coupled by Josephson coupling and we essentially deal with a large skyrmion, i.e. a bound state of fractional vortices but not two independent sublattices. I.e. in an absence of external field and away from system boundaries it will remain a bound state of vortices similar to that shown in Fig.\ref{Fig:charge12} }

{  Finally we comment on the comparison of these solutions to other cases of fractional, non-axially-symmetric vortices 
and Skyrmions (see e.g. \cite{volovik2003universe,knigavko1999magnetic,eto2013vortex,Garaud.Carlstrom.ea:13,agterberg2014microscopic,zyuzin2017nematic,silaev2015lifshitz}).
In contrast to superfluids, in isotropic superconductors the electromagnetic coupling and Josephson coupling strongly disfavor formation of Skyrmions and fractional vortices (see detailed discussion in e.g. \cite{frac}.  
Previously Skyrmions have been identified in several model superconducting systems, which include  either the terms that counter-act the Josephson and electromagnetic interactions (such as strong density-density interaction) or rely on complex interplay with other topological excitations coming from higher broken symmetry (such as fractionalization of vortices pinned by domain walls). By contrast the anistropic systems provide a new mechanism for Skyrmion formation that comes from hybridization of magnetic and Leggett modes leading to Josephson and electromagnetic coupling favouring Skyrmion formation, in contrast to their role in isotropic case.
}

\section{Conclusions}
In conclusion, in the most common s-wave case of multiband superconductors, in the bulk samples in the absence of thermal fluctuations, the vorticity in different bands is confined into a composite vortex due to intercomponent Josephson and electromagnetic coupling. The recently demonstrated hybridization of Leggett and London modes in the anisotropic case \cite{silaev2017non} raised the question of if the system can instead favour de-confinement of fractional vortices by favouring inter-band phase difference gradients when magnetic field is present.

We have demonstrated that unconventional electrodynamics of anisotropic multicomponent superconductors indeed leads, under certain conditions, to a change of the topological structure of vortex excitations. Namely, when anisotropy is sufficiently strong, the fractional vortices in the various bands repel each other in different ways (depending on the band). This leads to the constituents of an integer flux vortex becoming misaligned for $N>1$. As a result vortex solutions become unstable and the system forms extended textures, characterized by a nontrivial Skyrmionic charge. Importantly the Skyrmions form complicated bound states, which have a very rich structure and are characterized by two integers: number of magnetic flux quanta $N$ (equivalent to the winding of the phase sum of the superconducting components) and Skyrmionic charge $Q$. In general these two integers are not equal.

The Skyrmion formation is rather generic at sufficiently strong anisotropies: i.e. it does not require a  particular symmetry of the anisotropy and the effect exists also in the presence of additional components. 
We studied also a magnetization of a sample in an externally applied magnetic field: we found this leads to the formation of large-topological-charge Skyrmons extending through the system.

The Skyrmions found here are structurally very different from vortex solutions in the regimes considered in\cite{silaev2017non,winyard2017}. One of the key difference is that they are extended textures that do not have zeros of the total superfluid density. Although the Skyrmion formation required relatively strong anisotropies, the effect could also be present locally when there is strain in a sample.
 
 \begin{acknowledgments}
\noindent The work was supported by the Swedish Research Council Grants No. 642-2013-7837, VR2016-06122, Goran Gustafsson Foundation for Research in Natural Sciences and Medicine and EPSERC Grant No. EP/P024688/1. Part of the work was completed at the Aspen Center for Physics, which is supported by National Science Foundation grant PHY-1607611. The computations were performed on resources provided by the Swedish National Infrastructure for Computing (SNIC) at National Supercomputer 
Center at Link\"oping, Sweden. 
\end{acknowledgments}

\bibliography{references}
 \end{document}